\documentclass[aps,prb,reprint,amsmath,amssymb,amsfonts,showkeys,showpacs,preprintnumbers,floatfix]{revtex4-1}
\usepackage{bm,graphicx,xcolor,lineno,hyperref,microtype,multirow}
\newcommand{\eps}{\varepsilon}
\newcommand{\lam}{\lambda}
\newcommand{\mc}{\multicolumn}
\newcommand{\mr}{\multirow}
\newcommand{\Ec}{E_{\rm c}}
\newcommand{\ku}{k_{\uparrow}}
\newcommand{\kd}{k_{\downarrow}}

\newcommand{\upsa}{\Upsilon_0^{\text{a}}}
\newcommand{\upsb}{\Upsilon_0^{\text{b}}}
\newcommand{\Lama}{\Lambda_1^{\text{a}}}
\newcommand{\Lamb}{\Lambda_1^{\text{b}}}
\DeclareMathOperator{\Li}{Li}

\begin{document}

\title{Correlation energy of the spin-polarized uniform electron gas at high density}

\author{Pierre-Fran\c{c}ois Loos}
\email{loos@rsc.anu.edu.au}
\affiliation{Research School of Chemistry, 
Australian National University, Canberra, ACT 0200, Australia}
\author{Peter M. W. Gill}
\thanks{Corresponding author}
\email{peter.gill@anu.edu.au}
\affiliation{Research School of Chemistry, 
Australian National University, Canberra, ACT 0200, Australia}
\date{\today}

\begin{abstract}
The correlation energy per electron in the high-density uniform electron gas can be written as $\Ec(r_s,\zeta) = \lam_0(\zeta) \ln r_s + \eps_0(\zeta) + \lam_1(\zeta) \,r_s \ln r_s + O(r_s)$, where $r_s$ is the Seitz radius and $\zeta$ is the relative spin polarization.  We derive an expression for $\lam_1(\zeta)$ which is exact for any $\zeta$, including the paramagnetic and ferromagnetic limits, $\lam_1(0)$ and $\lam_1(1)$, and discover that the previously published $\lam_1(1)$ value is incorrect.  We trace this error to an integration and limit that do not commute.  The spin-resolution of $\lam_1(\zeta)$ into contributions of electron pairs is also derived.
\end{abstract}

\maketitle

The final decades of the twentieth century witnessed a major revolution in solid-state and molecular physics, as the introduction of sophisticated exchange-correlation models \cite{ParrYang} propelled density-functional theory (DFT) from qualitative to quantitative usefulness.  In principle, the foundation of DFT is the Hohenberg-Kohn theorem \cite{Hohenberg64} but, in practice, it is usually the supposed similarity between the electron density in a real system and the electron density in the hypothetical uniform electron gas (UEG). \cite{Vignale}

The three-dimensional UEG is characterized by a density $\rho = \rho_{\uparrow} + \rho_{\downarrow}$, where ${\rho_\uparrow}$ and ${\rho_\downarrow}$ is the (uniform) density of the spin-up and spin-down electrons, respectively.  In order to guarantee its stability, the electrons are assumed to be embedded in a uniform background of positive charge.

In 1965, Kohn and Sham \cite{Kohn65} showed that the knowledge of a analytical parametrization of the UEG correlation energy allows one to perform approximate calculations for atoms, molecules and solids.  This led to the development of various spin-density correlation functionals (VWN, \cite{Vosko80} PZ, \cite{Perdew81} PW92, \cite{Perdew92} \textit{etc.}), all of which require information on the high- and low-density regimes of the spin-polarized UEG, and are parametrized using results from near-exact Quantum Monte Carlo (QMC) calculations. \cite{Ceperley80, Ballone92, Ortiz94, Ortiz97, Kwon98, Ortiz99, Zong02, Zhang08}

However, inspired by Wigner's seminal work, \cite{Wigner32} Sun, Perdew and Seidl have recently shown that the correlation energy of the UEG can be estimated accurately without any QMC input. \cite{Sun10}  They used a ``density-parameter interpolation'' (DPI) between the (near-)exact high- and low-density regimes, which precisely reproduces the first few coefficients of the high- and low-density energy expansions. \footnote{The approximate expression of the correlation energy in Ref. \onlinecite{Sun10} is not an interpolation formula which uses QMC data, unlike the well-known PZ \cite{Perdew81} and PW92 \cite{Perdew92} functionals.  This interpolation formula uses the exact or near-exact coefficients of the high- and low-density limits embedded in a Pad\'e-like approximant.  The only QMC information used in Ref. \onlinecite{Sun10} is the value of the correlation energy at the ferromagnetic transition, \cite{Ceperley80} which is required to build an approximate expression of the fourth-order coefficient ($r_s$ term).}  Knowledge of these coefficients, of course, is essential for such interpolations and is the motivation for the present work.

The high-density expansion of the correlation energy per electron (or reduced energy) of the UEG is \cite{Wigner32, Macke50, Bohm53, Pines53, GellMann57, DuBois59, Carr64, Misawa65, Onsager66, Wang91, Hoffman92, Sun10} 
\begin{equation}
\label{Ecjellium}
        \Ec(r_s,\zeta) = \lam_0(\zeta) \ln r_s + \eps_0(\zeta) + \lam_1(\zeta) \,r_s \ln r_s + O(r_s),
\end{equation}
where $r_s = \left(4\pi\rho/3\right)^{-1/3}$ is the so-called Seitz radius, and $\zeta = (\rho_{\uparrow} - \rho_{\downarrow})/\rho$ is the relative spin polarization.  It is clear that $\lam_0(\zeta), \,\eps_0(\zeta), \,\lam_1(\zeta), \,\ldots$ must be even functions.  We use atomic units throughout.

The coefficient $\lam_0(\zeta)$ can be obtained by the Gell-Mann--Brueckner resummation technique, \cite{GellMann57} which sums the most divergent terms of the series \eqref{Ecjellium} to obtain
\begin{equation}
	\lam_0(\zeta) = \frac{3}{32\pi^3} \int_{-\infty}^{\infty} \left[R_0(u,\zeta)\right]^2 du,
\end{equation}
where
\begin{gather}
	\label{R0u-zeta}
	R_0(u,\zeta) = \kd R_0\left(\frac{u}{\kd}\right) + \ku R_0\left(\frac{u}{\ku}\right),
	\\
	\label{R0u}
	R_0(u) = 1 - u \arctan(1/u),
\end{gather}
and $k_{\uparrow,\downarrow} = \left(1\pm\zeta\right)^{1/3}$ is the Fermi momentum of the spin-up or spin-down electrons.  The paramagnetic \cite{Macke50} ($\zeta=0$) and ferromagnetic \cite{Misawa65} ($\zeta=1$) limits are given in Table \ref{tab:coeffs} and the spin-scaling function
\begin{align} \label{Lam0-zeta}
	\Lambda_0(\zeta)	& = \frac{\lam_0(\zeta)}{\lam_0(0)}							\notag	\\
						& = \frac{1}{2} + \frac{1}{4(1-\ln 2)} \bigg[\kd \ku (\kd+\ku)	\notag	\\
						& \quad - \kd^3 \ln \left(1+\frac{\ku}{\kd}\right) - \ku^3 \ln \left(1+\frac{\kd}{\ku}\right) \bigg]
\end{align}
was obtained by Wang and Perdew. \cite{Wang91}

\begin{table*}
\caption{
\label{tab:coeffs}
Energy coefficients and spin-scaling functions of the paramagnetic ($\zeta=0$) and ferromagnetic ($\zeta=1$) states of the high-density UEG.  Note that $\alpha = \left(9\pi/4\right)^{-1/3}$ and $z(n)$ is the Riemann zeta function. \cite{NISTbook}}
\begin{ruledtabular}
\begin{tabular}{ccccc}
	\mr{2}{*}{Term}		&	\mr{2}{*}{Coefficient}		&	Paramagnetic limit		&	Ferromagnetic limit		&	Spin-scaling function
\\
				&					&	$\eps(0)$, $\lam(0)$		&	$\eps(1)$, $\lam(1)$		&	$\Upsilon(\zeta)$, $\Lambda(\zeta)$		
\\
\hline		
\\
	$\ln r_s$		&	$\lam_{0}(\zeta)$		&	$\displaystyle \frac{1-\ln2}{\pi^2}$
									&	$\displaystyle \frac{1-\ln2}{2\pi^2}$					&	Eq.~\eqref{Lam0-zeta}
\\[10pt]
	\mr{2}{*}{$r_s^0$}	&	$\eps_0^{\text{a}}(\zeta)$	&	$\displaystyle -0.0710995$
									&	$\displaystyle -0.0499167$						&	Ref.~\onlinecite{Hoffman92}
\\[10pt]
				&	$\eps_0^{\text{b}}(\zeta)$	&	$\displaystyle \frac{\ln 2}{6} - \frac{3}{4\pi^2} z(3)$
									&	$\displaystyle \frac{\ln 2}{6} - \frac{3}{4\pi^2} z(3)$		&	1
\\[10pt]
	\mr{2}{*}{$r_s\ln r_s$}	&	$\lam_1^{\text{a}}(\zeta)$	&	$\displaystyle \frac{\alpha}{24\pi^3}(\pi^2-6)$
									&	$\displaystyle \frac{1}{2^{7/3}} \frac{\alpha}{24\pi^3}(\pi^2+6)$
																			&	Eq.~\eqref{Lama-zeta}
\\[10pt]
				&	$\lam_1^{\text{b}}(\zeta)$	&	$\displaystyle \frac{\alpha}{4\pi^3} (\pi^2-12\ln 2)$
									&	$\displaystyle \frac{1}{2^{4/3}} \frac{\alpha}{4\pi^3} (\pi^2-12\ln 2)$
																			&	Eq.~\eqref{Lamb-zeta}
\\[10pt]
\end{tabular}
\end{ruledtabular}
\end{table*}

\begin{table}
\caption{
\label{tab:Ec1}
Reduced correlation energy $-\Ec(r_s,1)$ for the ferromagnetic state of the UEG for various $r_s$.}
\begin{ruledtabular}
\begin{tabular}{cllllll}
	$r_s$	&	\mc{1}{c}{QMC\footnotemark[1]}		&	\mc{1}{c}{DPI\footnotemark[2]}	&	\mc{1}{c}{Modified-DPI\footnotemark[3]}	
\\                                      
\hline 
	$2$	&	0.0240(3)	&	0.0236		&	0.0238		
\\                                      
	$5$	&	0.0154(1)	&	0.0151		&	0.0152		
\\                                      
	$10$	&	0.0105(1)	&	0.0102		&	0.0103		
\\                                      
	$20$	&	0.00678(2)	&	0.00663		&	0.00664		
\\                                      
	$50$	&	0.00355(1)	&	0.00350		&	0.00350		
\\                                      
	$100$	&	0.002073(3)	&	0.002055	&	0.002055	
\end{tabular}
\end{ruledtabular}
\footnotetext[1]{Benchmark QMC results taken from Ref.~\onlinecite{Ceperley80}.  The digits in parenthesis represent the error bar in the last decimal place.}
\footnotetext[2]{Results taken from Ref.~\onlinecite{Sun10} using the DPI (density-parameter interpolation) formula with $\lam_1(1) = 0.003125$.}
\footnotetext[3]{Results from the present work using the DPI formula with $\lam_1(1) = 0.004792$.}
\end{table}

\begin{figure}
	\includegraphics[width=0.4\textwidth]{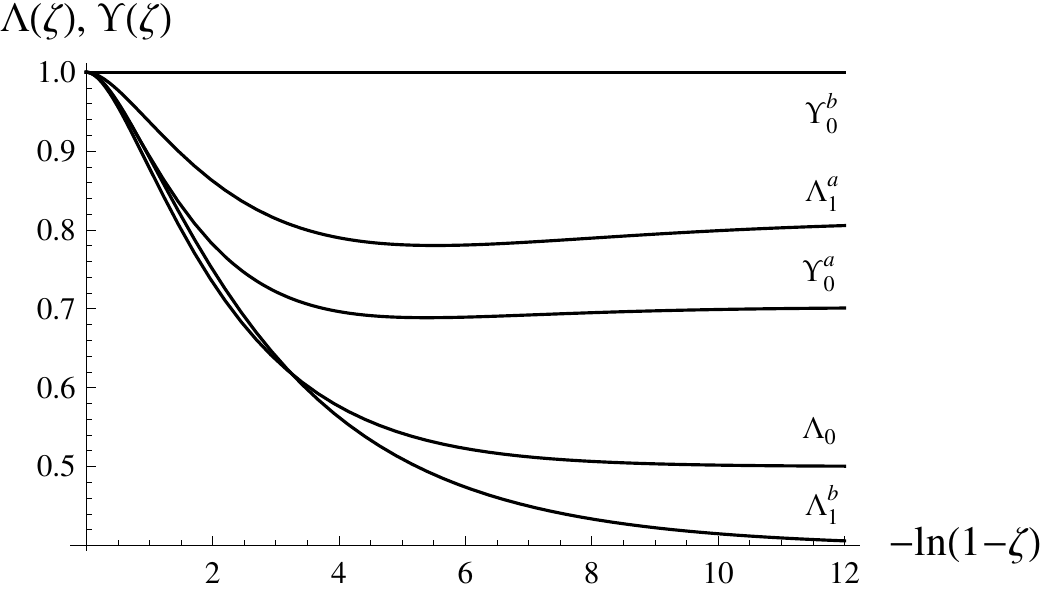}
	\caption{
	\label{fig:spin-scaling}
	The five spin-scalings as functions $\zeta$.
	}
\end{figure}

The coefficient $\eps_0(\zeta)$ is usually written as the sum
\begin{equation} \label{eps0}
	\eps_0(\zeta) = \eps_{0}^{\text{a}}(\zeta) + \eps_{0}^{\text{b}}
\end{equation}
of a RPA (random-phase approximation) term $\eps_{0}^{\text{a}}(\zeta)$ and a first-order exchange term $\eps_0^{\text{b}}$.  The RPA term $\eps_{0}^{\text{a}}(\zeta)$ is not known in closed form, but it can be computed numerically with high precision. \cite{Hoffman92}  Its paramagnetic and ferromagnetic limits are given in Table \ref{tab:coeffs} and the spin-scaling function $\upsa(\zeta) = \eps_0^{\text{a}}(\zeta)/\eps_0^{\text{a}}(0)$ can be found using Eq.~(20) in Ref.~\onlinecite{Hoffman92}.  The first-order exchange term \cite{Onsager66} is given in Table \ref{tab:coeffs} and, because it is independent of the spin-polarization, the spin-scaling function $\upsb(\zeta) = \eps_0^{\text{b}}(\zeta)/\eps_0^{\text{b}}(0) = 1$ is trivial.

The coefficient $\lam_1(\zeta)$ can be written similarly \cite{Carr64} as 
\begin{equation}
	\lam_1(\zeta) = \lam_{1}^{\text{a}}(\zeta) + \lam_{1}^{\text{b}}(\zeta),
\end{equation}
where
\begin{gather}
	\lam_{1}^{\text{a}}(\zeta) = -\frac{3\alpha}{8\pi^5} \int_{-\infty}^{\infty} \mathcal{R}_1^{\text{a}}(u,\zeta) \,du,	
	\label{lam1a-zeta}	
	\\
	\lam_{1}^{\text{b}}(\zeta) = \frac{3\alpha}{16\pi^4} \int_{-\infty}^{\infty} \mathcal{R}_1^{\text{b}}(u,\zeta) \,du	
	\label{lam1b-zeta}
\end{gather}
are the RPA and second-order exchange contributions and $\alpha = \left(9\pi/4\right)^{-1/3}$.  The integrand functions are \cite{Perdew92, Sun10}
\begin{gather}
	\mathcal{R}_1^{\text{a}}(u,\zeta) = R_0(u,\zeta)^2 R_1(u,\zeta),	\\
	\mathcal{R}_1^{\text{b}}(u,\zeta) = R_0(u,\zeta) R_2(i u,\zeta),	\\
	\label{R1u-zeta}
	R_1(u,\zeta) = \kd^{-1} R_1\left(\frac{u}{\kd}\right) + \ku^{-1} R_1\left(\frac{u}{\ku}\right),	
	\\
	R_2(i u,\zeta) = R_2\left(i\frac{u}{\kd}\right) + R_2\left(i\frac{u}{\ku}\right),
	\\
	R_1(u) = -\frac{\pi}{3(1+u^2)^2},					\\
	R_2(i u) = 4\frac{(1+3u^2) - u(2+3u^2) \arctan u}{1+u^2}.
\end{gather}
Carr and Maradudin gave an estimate \cite{Carr64} of $\lam_1(0)$ and this was later refined by Perdew and coworkers. \cite{Perdew92, Sun10}

However, we have found that the integrals in Eqs.~\ref{lam1a-zeta} and \ref{lam1b-zeta} can be evaluated exactly by computer software, \cite{Mathematica7} giving the paramagnetic and ferromagnetic values in Table \ref{tab:coeffs} and the spin-scaling functions
\begin{widetext}
	\begin{multline} 
		\label{Lama-zeta}
		\Lama(\zeta) = \frac{3}{\pi^2-6} \left\{ \left(\frac{\pi^2}{6}+\frac{1}{4}\right)(\kd^2+\ku^2)-\frac{3}{2}\kd\ku
				-\frac{\kd^2+\ku^2}{\kd^2-\ku^2}\kd\ku\ln\left(\frac{\kd}{\ku}\right) \right. \\
			\left. -\frac{\kd^2-\ku^2}{2}\left[\Li_2\left(\frac{\kd-\ku}{\kd+\ku}\right) - \Li_2\left(\frac{\ku-\kd}{\kd+\ku}\right) \right]\right\},
	\end{multline}
	\begin{multline} 
		\label{Lamb-zeta}
		\Lamb(\zeta) = \frac{3}{\pi^2-12\ln2} \left\{ \frac{\pi^2}{6}(\kd^2+\ku^2) + (1-\ln2)(\kd-\ku)^2
				- \frac{\kd^2}{2}\Li_2\left(\frac{\kd-\ku}{\kd+\ku}\right) - \frac{\ku^2}{2}\Li_2\left(\frac{\ku-\kd}{\kd+\ku}\right)	\right.	\\
			\left. + \frac{1}{\kd\ku} \left[\kd^4\ln\left(\frac{\kd}{\kd+\ku}\right) + \kd^2\ku^2\ln\left(\frac{\kd\ku}{(\kd+\ku)^2}\right)
				+ \ku^4\ln\left(\frac{\ku}{\kd+\ku}\right) \right] \right\},
	\end{multline}
\end{widetext}
where $\Li_2$ is the dilogarithm function. \cite{NISTbook}

The spin-scalings $\Lambda_0(\zeta)$, $\upsa(\zeta)$, $\upsb(\zeta)$, $\Lama(\zeta)$ and $\Lamb(\zeta)$ are shown in Fig.~\ref{fig:spin-scaling}, highlighting the Hoffmann minimum \cite{Hoffman92} in $\upsa(\zeta)$ near $\zeta=0.9956$ and revealing a similar minimum in $\Lama(\zeta)$ near $\zeta= 0.9960$.  Such minima seem to be ubiquitous in RPA coefficients.

The data in Table \ref{tab:coeffs} yield the exact values
\begin{align}
	\lam_1(0)	& = \frac{\alpha}{4\pi^3} \left(\frac{7\pi^2}{6} - 12 \ln 2 - 1\right)						\notag			\\
			& = 0.00922921\ldots,																\label{lam1-0}	\\
	\lam_1(1)	& = 2^{-4/3} \frac{\alpha}{4\pi^3} \left( \frac{13\pi^2}{12} -12\ln 2 +\frac{1}{2} \right)	\notag			\\
			& = 0.00479225\ldots,																\label{lam1-1}
\end{align}
and it is revealing to compare these with recent numerical calculations.  The estimate $\lam_1(0) \approx 0.0092292$ by Sun \textit{et al.} \cite{Sun10} agrees perfectly with Eq.~\eqref{lam1-0} but their estimate $\lam_1(1) \approx 0.003125$ is strikingly different from Eq.~\eqref{lam1-1}.  How can this discrepancy be explained?

Following Gell-Mann and Brueckner, \cite{GellMann57} and Ueda, \cite{Ueda61} Misawa argued \cite{Misawa65} that the $\zeta=0$ and $\zeta=1$ limits of the RPA and exchange contributions to the correlation energy are related by
\begin{align}
	\Ec^{\text{a}}(r_s,1) & = \frac{1}{2} \Ec^{\text{a}}(2^{-4/3} r_s,0),	\\
	\Ec^{\text{b}}(r_s,1) & = \Ec^{\text{b}}(2^{-4/3}r_s,0),
\end{align}
and, from these relations, Perdew and Wang inferred \cite{Perdew92}
\begin{align}
	\lam_1^{\text{a}}(1) & = 2^{-7/3} \lam_1^{\text{a}}(0),	\label{lam1a-1-PW}	\\
	\lam_1^{\text{b}}(1) & = 2^{-4/3} \lam_1^{\text{b}}(0).	\label{lam1b-1-PW}
\end{align}
These are also obtained if the $\zeta \to 1$ limit of the integrands in Eqs.~\ref{lam1a-zeta} and \ref{lam1b-zeta} is taken before integrating over $u$.

Numerical evaluations of Eq.~\eqref{lam1b-zeta} and analytical results from Eq.~\eqref{Lamb-zeta} confirm that Eq.~\eqref{lam1b-1-PW} is correct.  However, numerical evaluations of Eq.~\ref{lam1a-zeta} and analytical results from Eq.~\eqref{Lama-zeta} agree that Eq.~\eqref{lam1a-1-PW} is wrong and that, in fact,
\begin{equation}
	\lam_1^{\text{a}}(1) = 2^{-7/3} \lam_{1}^{\text{a}}(0) \times \frac{\pi^2+6}{\pi^2-6}.
\end{equation}
The error in Eq.~\eqref{lam1a-1-PW} arises from the non-commutivity of the $\zeta\to1$ limit and the $u$ integration, which is due to the non-uniform convergence of $\mathcal{R}_1^{\text{a}}(u,\zeta)$.

To show this particular point, let us define
\begin{gather}
	\label{DL}
	\Delta\lam_{1}^{\text{a}}(\zeta)
	= -\frac{3\alpha}{8\pi^5} \int_{-\infty}^{\infty} \Delta \mathcal{R}_1^{\text{a}}(u,\zeta) \,du,
	\\
	\label{DR}
	\Delta \mathcal{R}_{1}^{\text{a}}(u,\zeta) 
	= \frac{\ku^2}{\kd} R_0\left(\frac{u}{\ku}\right)^2 R_1\left(\frac{u}{\kd}\right).
\end{gather}
It can be easily shown that it is not possible to find a function $D(u)$, which is integrable with respect to $u$ and dominates $\Delta \mathcal{R}_{1}^{\text{a}}(u,\zeta)$, {\em i.e.} $\forall (u,\zeta),\: \left|\Delta \mathcal{R}_{1}^{\text{a}}(u,\zeta) \right| \le D(u)$.   Thus, according to the dominated convergence theorem, one cannot show that limit and integration can be interchanged, and 
\begin{align}
	\lim_{\zeta\to1}\Delta\mathcal{R}_{1}^{\text{a}}(u,\zeta) & = 0,
	& 
	\Delta\lam_{1}^{\text{a}}(1) & = 0.
\end{align}

However, substituting $t=u/\kd$ in Eqs.~\eqref{DL} and \eqref{DR}, one immediately finds a function $D(t)$, which is integrable with respect to $t$ and dominates $\Delta \mathcal{R}_{1}^{\text{a}}(t,\zeta)$.  This ensures the possibility of interchanging limit and integration. It yields
\begin{gather}
	\lim_{\zeta\to1}\Delta\mathcal{R}_{1}^{\text{a}}(t,\zeta) = - \frac{2^{2/3} \pi}{3(1+t^2)^2},
	\\
	\label{DL-res}
	\begin{split}
	\Delta\lam_{1}^{\text{a}}(1) 
	& = \frac{3\alpha}{8\pi^5} \int_{-\infty}^{\infty} \frac{2^{2/3} \pi}{3(1+t^2)^2}\,dt 
	\\
	& = 2^{-1/3} \frac{\alpha}{8\pi^3} = 0.00166727,
	\end{split}
\end{gather}
which is exactly the difference between the two values of $\lam_{1}^{\text{a}}(1)$.

The effect of the coefficient $\lam_1(1)$ on the reduced correlation energy of the ferromagnetic state has been studied by varying its value in the DPI formula proposed by Sun {\em et al.} in Ref.~\onlinecite{Sun10}. The results have been compared with the benchmark QMC calculations of Ceperley and Alder. \cite{Ceperley80}  As shown in Table \ref{tab:Ec1}, the new value of $\lam_1(1)$ derived in the present study systematically improves the accuracy of the DPI correlation energy, especially for small $r_s$.

\begin{figure*}
\begin{tabular}{ccc}
	\includegraphics[width=0.28\textwidth]{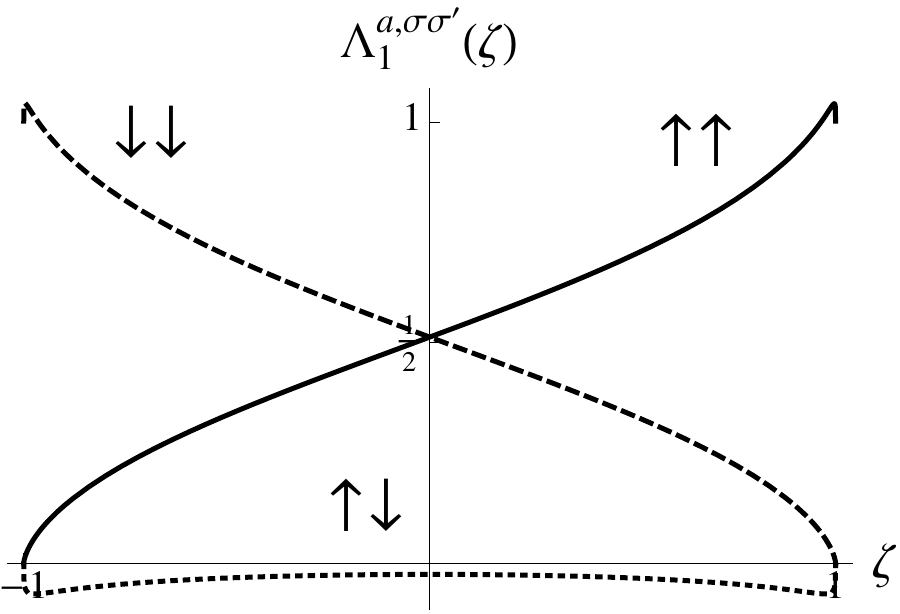}
	&
	\includegraphics[width=0.28\textwidth]{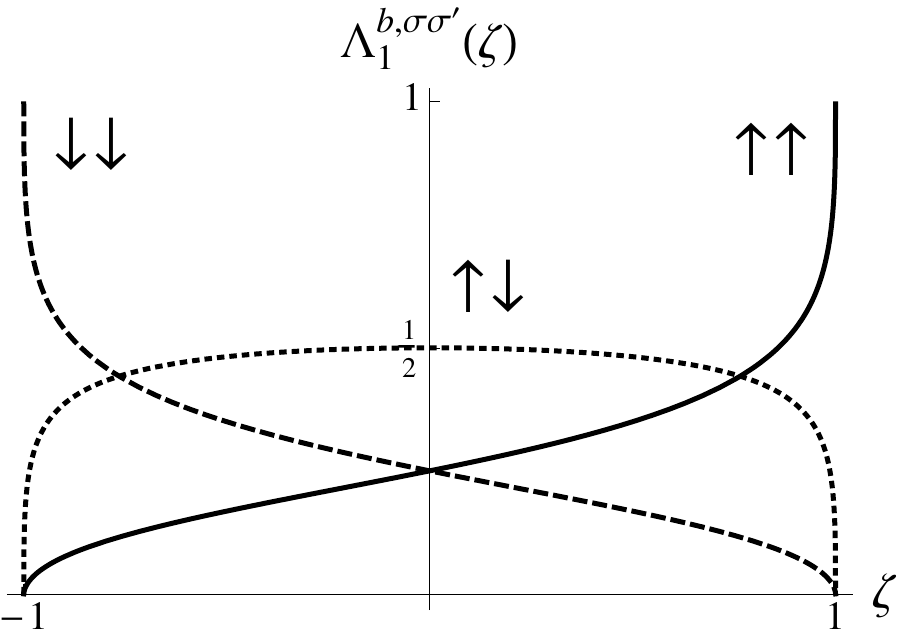}
	&
	\includegraphics[width=0.28\textwidth]{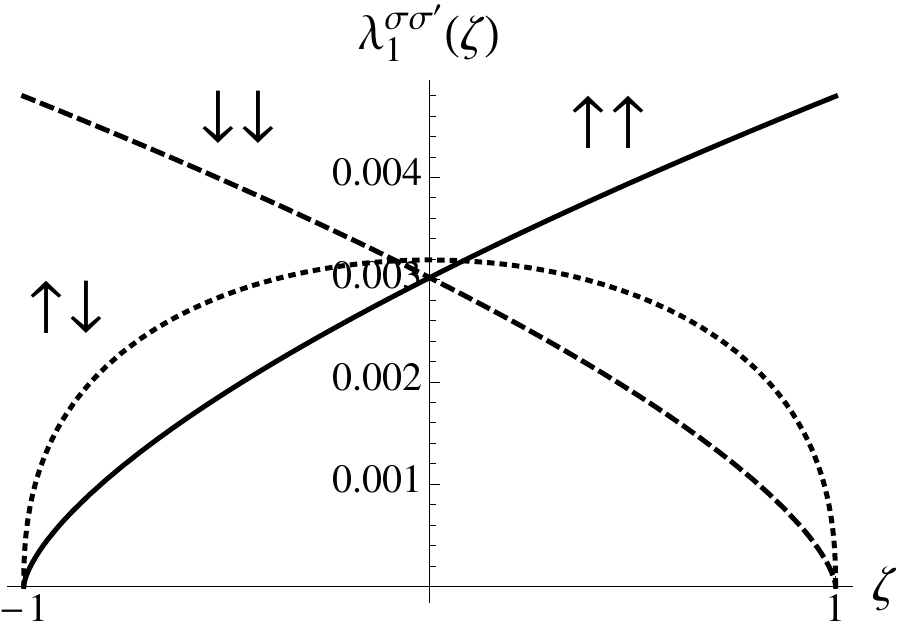}
\end{tabular}
	\caption{
	\label{fig:spin-res}
	Spin-resolution of $\Lambda_{1}^{\text{a},\sigma\sigma^{\prime}}(\zeta)$, $\Lambda_{1}^{\text{b},\sigma\sigma^{\prime}}(\zeta)$ and $\lambda_{1}^{\sigma\sigma^{\prime}}(\zeta)$ as functions of $\zeta$.
	}
\end{figure*}

In some cases, \cite{Wang91, GoriGiorgi04} it is of interest to resolve $\lambda_{1}(\zeta)$ into contributions due to $\uparrow\uparrow$, $\downarrow\downarrow$ and $\uparrow\downarrow$ electron pairs, such as
\begin{gather}
	\lambda_{1}^{\text{i}}(\zeta) 
	= \lambda_{1}^{\text{i},\uparrow\uparrow}(\zeta) 
	+ \lambda_{1}^{\text{i},\downarrow\downarrow}(\zeta) 
	+ \lambda_{1}^{\text{i},\uparrow\downarrow}(\zeta),
	\\
	\Lambda_{1}^{\text{i},\sigma\sigma^{\prime}}(\zeta) 
	= \frac{\lambda_{1}^{\text{i},\sigma\sigma^{\prime}}(\zeta)}{\lambda_{1}^{\text{i}}(\zeta)},
\end{gather}
where $\text{i} = \text{a or b}$, and $\sigma\sigma^{\prime} =\;\uparrow\uparrow,\;\downarrow\downarrow \text{or} \uparrow\downarrow$.  Using \eqref{Lama-zeta} and \eqref{Lamb-zeta}, we find
\begin{align}
	\Lambda_{1}^{\text{a},\uparrow\uparrow}(\zeta)
	& = \frac{1}{8} \frac{\pi^2+6}{\pi^2-6}\frac{(1+\zeta)^{2/3}}{\Lambda_{1}^{\text{a}}(\zeta)},
	\\
	\Lambda_{1}^{\text{b},\uparrow\uparrow}(\zeta)
	& = \frac{1}{4} \frac{(1+\zeta)^{2/3}}{\Lambda_{1}^{\text{b}}(\zeta)}.
\end{align}
The remaining contributions can be obtained using the relations
\begin{align}
	\Lambda_{1}^{\text{i},\downarrow\downarrow}(\zeta) 
	& = \Lambda_{1}^{\text{i},\uparrow\uparrow}(-\zeta),
	\\
	\Lambda_{1}^{\text{i},\uparrow\downarrow}(\zeta) 
	& = 1 - \Lambda_{1}^{\text{i},\uparrow\uparrow}(\zeta) - \Lambda_{1}^{\text{i},\downarrow\downarrow}(\zeta),
\end{align}
and are represented in Fig.~\ref{fig:spin-res}.

In conclusion, we have found a closed-form expression for the coefficient $\lam_1(\zeta)$ of the $r_s \ln r_s$ term in Eq.~\eqref{Ecjellium}.  It is valid for any value of $\zeta$ and, in particular, for the paramagnetic ($\zeta=0$) and ferromagnetic ($\zeta=1$) limits.  This reveals that an earlier derivation of the ferromagnetic limit $\lam_1(1)$ was incorrect because of an inadmissible interchange of a limit and an integral.  The present result has no direct impact on the quantum phase diagram of the UEG, because the effect of the coefficient $\lambda_1(\zeta)$ is more pronounced in the high-density limit ($0<r_s\lesssim 2$), where the paramagnetic fluid is significantly more stable than the ferromagnetic one. \cite{Ceperley80}  Preliminary results on higher-order coefficients reveal that they behave similarly, and special care has to be taken in future studies.  We believe that these new results will be useful in the future development of exchange-correlation functionals within DFT.  

The authors thank an anonymous referee for providing helpful comments leading to Eqs.~\eqref{DL}--\eqref{DL-res}.  P.M.W.G. thanks the NCI National Facility for a generous grant of supercomputer time and the Australian Research Council (Grants DP0984806 and DP1094170) for funding.

\end{document}